\documentclass[aps,showpacs,twocolumn,superscriptaddress]{revtex4}
\usepackage{amssymb}
\usepackage{epsfig}
\usepackage{graphicx,amsmath}
\begin{document}

\title{Energy scales and the non-Fermi liquid behavior in $\rm \bf YbRh_2Si_2$}
\author{V.R. Shaginyan}\email{vrshag@thd.pnpi.spb.ru}
\affiliation{Petersburg Nuclear Physics Institute, RAS, Gatchina,
188300, Russia}\affiliation{Racah Institute of Physics, Hebrew
University, Jerusalem 91904, Israel}
\author{M.Ya. Amusia}\affiliation{Racah Institute
of Physics, Hebrew University, Jerusalem 91904, Israel}
\author{K.G. Popov}\affiliation{Komi Science Center, Ural Division,
RAS, Syktyvkar, 167982, Russia} \author{S.A. Artamonov}
\affiliation{Petersburg Nuclear Physics Institute, RAS, Gatchina,
188300, Russia}

\begin{abstract}

Multiple energy scales are detected in measurements of the
thermodynamic and transport properties in heavy fermion metals. We
demonstrate that the experimental data on the energy scales can be
well described by the scaling behavior of the effective mass at the
fermion condensation quantum phase transition, and show that the
dependence of the effective mass on temperature and applied magnetic
fields gives rise to the non-Fermi liquid behavior. Our analysis is
placed in the context of recent salient experimental results. Our
calculations of the non-Fermi liquid behavior, of the scales and
thermodynamic and transport properties are in good agreement with
the heat capacity, magnetization, longitudinal magnetoresistance and
magnetic entropy obtained in remarkable measurements on the heavy
fermion metal $\rm YbRh_2Si_2$.
\end{abstract}

 \maketitle

An explanation of the rich and striking behavior of strongly
correlated electron ensemble in heavy fermion (HF) metals in the
vicinity of a quantum phase transition is, as years before, among
the main problems  of the condensed matter physics. It is common
wisdom that low-temperature and quantum fluctuations at quantum
phase transitions form the specific heat, magnetization,
magnetoresistance etc. which are drastically different from those of
ordinary metals \cite{senth,col,lohneysen,si,sach}. Conventional
arguments that quasiparticles in strongly correlated Fermi liquids
"get heavy and die" at the QCP commonly employ the well-known
formula basing on assumptions that the $z$-factor (the quasiparticle
weight in the single-particle state) vanishes at the points of
second-order phase transitions \cite{col1}. However, it has been
shown this scenario is problematic \cite{khodz}. On the other hand,
facts collected on HF metals demonstrate that the effective mass
strongly depends on temperature $T$, doping (or the number density)
$x$ and applied magnetic fields $B$, while the effective mass $M^*$
itself can reach very high values or even diverge, see e.g.
\cite{lohneysen,si}. Such a behavior is so unusual that the
traditional Landau quasiparticles paradigm does not apply to it. The
paradigm says that elementary excitations determine the physics at
low temperatures. These behave as Fermi quasiparticles and have a
certain effective mass $M^*$ which is independent of $T$, $x$, and
$B$ and is a parameter of the theory \cite{land}.

A concept of fermion condensation quantum phase transition (FCQPT)
preserving quasiparticles and intimately related to the unlimited
growth of $M^*$, had been suggested  \cite{khs,ams,volovik}. Studies
show that it is capable to deliver an adequate theoretical
explanation of vast majority of experimental results in different HF
metals \cite{obz,khodb}. In contrast to the Landau paradigm based on
the assumption that $M^*$ is a constant, in FCQPT approach $M^*$
strongly depends on $T$, $x$, $B$ etc. Therefore, in accord with
numerous experimental facts the extended quasiparticles paradigm is
to be introduced. The main point here is that the well-defined
quasiparticles determine as before the thermodynamic and transport
properties of strongly correlated Fermi-systems, $M^*$ becomes a
function of $T$, $x$, $B$, while the dependence of the effective
mass on $T$, $x$, $B$ gives rise to the non-Fermi liquid (NFL)
behavior \cite{obz,khodb,zph,ckz,plaq}.

In this letter, we analyze the NFL behavior of strongly correlated
Fermi systems and show that this is generated by the dependence of
the effective mass on temperature, number density and magnetic field
at FCQPT. We demonstrate that the NFL behavior observed in the
transport and thermodynamic properties of HF metals can be described
in terms of the scaling behavior of the normalized effective mass.
This allows us to construct the scaled thermodynamic and transport
properties extracted from experimental facts in wide range of the
variation of scaled variable. We show that "peculiar points" of the
normalized effective mass give rise to the energy scales observed in
the thermodynamic and transport properties of HF metals. Our
calculations of the thermodynamic and transport properties are in
good agreement with the heat capacity, magnetization, longitudinal
magnetoresistance and magnetic entropy obtained in remarkable
measurements on the heavy fermion metal $\rm YbRh_2Si_2$
\cite{steg1,oes,steg,geg}. For $\rm YbRh_2Si_2$ the constructed
thermodynamic and transport functions extracted from experimental
facts show the scaling over three decades in the variable.

To avoid difficulties associated with the anisotropy generated by
the crystal lattice of solids, we study the universal behavior of
heavy-fermion metals using the model of the homogeneous
heavy-electron (fermion) liquid \cite{plaq,shag1,shag2}.

We start with visualizing the main properties of FCQPT. To this end,
consider the density functional theory for superconductors (SCDFT)
\cite{gross}. SCDFT states that at fixed temperature $T$ the
thermodynamic potential $\Phi$ is a universal functional of the
number density $n({\bf r})$ and the anomalous density (or the order
parameter) $\kappa({\bf r},{\bf r}_1)$ and provides a variational
principle to determine the densities \cite{gross}. At the
superconducting transition temperature $T_c$ a superconducting state
undergoes the second order phase transition. Our goal now is to
construct a quantum phase transition which evolves from the
superconducting one. In that case, the superconducting state takes
place at $T=0$ while at finite temperatures there is a normal state.
This means that in this state the anomalous density is finite while
the superconducting gap vanishes. For the sake of simplicity, we
consider a homogeneous Fermi (electron) system. Then, the
thermodynamic potential reduces to the ground state energy $E$ which
turns out to be a functional of the occupation number $n({\bf p})$
since $\kappa=\sqrt{n(1-n)}$ \cite{dft,gross,yakov,plaq}. Upon
minimizing $E$ with respect to $n({\bf p})$, we obtain
\begin{equation}\label{FCM}
\frac{\delta E}{\delta n({\bf p})}=\varepsilon({\bf
p})=\mu,\end{equation} where $\mu$ is the chemical potential. It is
seen from Eq. \eqref{FCM} that instead of the Fermi step, we have
$0<n(p)<1$ in certain range of momenta $p_i\leq p\leq p_f$ with
$\kappa$ is finite in this range. Thus, the step-like Fermi filling
inevitably undergoes restructuring and formes the fermion condensate
(FC) as soon as Eq. \eqref{FCM} possesses not-trivial solutions at
some point $x=x_c$ when $p_i=p_f=p_F$ \cite{khs,obz,khodb}. Here
$p_F$ is the Fermi momentum and $x =p_F^3/3\pi^2$.

At any small but finite temperature the anomalous density $\kappa$
(or the order parameter) decays and this state undergoes the first
order phase transition and converts into a normal state
characterized by the thermodynamic potential $\Phi_0$. At $T\to0$,
the entropy $S=-\partial \Phi_0/\partial T$ of the normal state is
given by the well-known relation \cite{land}
\begin{equation}
S_0=-2\int[n({\bf p}) \ln (n({\bf p}))+(1-n({\bf p})\ln (1-n({\bf
p}))]\frac{d{\bf p}}{(2\pi) ^3},\label{SN}
\end{equation}
which follows from combinatorial reasoning. Since the entropy of the
superconducting ground state is zero, it follows from Eq. \eqref{SN}
that the entropy is discontinuous at the phase transition point,
with its discontinuity $\Delta S=S_0$. The latent heat $q$ of
transition from the asymmetrical to the symmetrical phase is
$q=T_cS_0=0$ since $T_c=0$. Because of the stability condition at
the point of the first order phase transition, we have
$\Phi_0[n({\bf p})]=\Phi[\kappa({\bf p})]$. Obviously the condition
is satisfied since $q=0$.

At $T=0$, a quantum phase transition is driven by a nonthermal
control parameter, e.g. the number density $x$. To clarify the role
of $x$, consider the effective mass $M^*$ which is related to the
bare electron mass $m$ by the well-known Landau equation \cite{land}
which is valid when $M^*$ strongly depends on $B$, $T$ or $x$
\cite{plaq}
\begin{equation}\label{LANDM}
\frac{1}{M^*}=\frac{1}{m}+\int \frac{{\bf p}_F{\bf p_1}}{p_F^3}
F({\bf p_F},{\bf p}_1)\frac{\partial n(p_1,T)}{\partial p_1}
\frac{d{\bf p}_1}{(2\pi)^3}.
\end{equation}
Here we omit the spin indices for simplicity, $n({\bf p},T)$ is
quasiparticle occupation number, and $F$ is the Landau amplitude. At
$T=0$, Eq. \eqref{LANDM} reads \cite{pfit,pfit1}
\begin{equation}\label{MM*}
\frac{M^*}{m}=\frac{1}{1-N_0F^1(x)/3}.\end{equation} Here $N_0$ is
the density of states of free electron gas and $F^1(x)$ is the
$p$-wave component of Landau interaction amplitude $F$. When at some
critical point $x=x_c$, $F^1(x)$ achieves certain threshold value,
the denominator in Eq. \eqref{MM*} tends to zero so that the
effective mass diverges at $T=0$ \cite{pfit,pfit1,shag3}. It follows
from Eq. \eqref{MM*} that beyond the quantum critical point (QCP)
$x_c$, the effective mass becomes negative. To avoid unstable and
physically meaningless state with a negative effective mass, the
system must undergo a quantum phase transition at QCP $x=x_c$
\cite{khs,ams,obz,khodb}.

\begin{figure} [! ht]
\begin{center}
\vspace*{-0.5cm}
\includegraphics [width=0.49\textwidth]{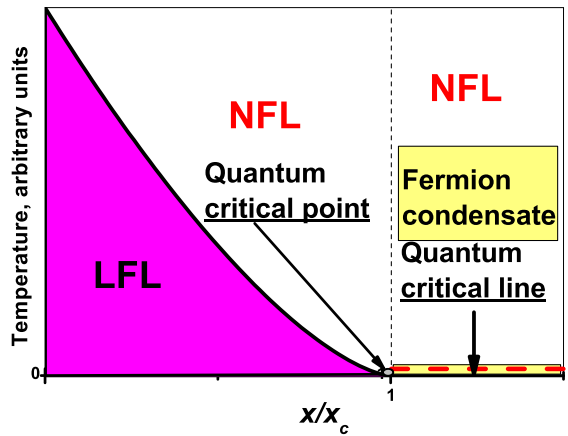}
\end{center}
\vspace*{-0.8cm} \caption{Schematic phase diagram of the system
driven to the FC state. The number density $x$ is taken as the
control parameter and depicted as $x/x_c$. The quantum critical
point (QCP), $x/x_c=1$, of FCQPT is shown by the arrow. At $x/x_c<1$
and sufficiently low temperatures, the system is in the Landau Fermi
liquid (LFL) state as shown by the shadow area. At $T=0$ and beyond
QCP, $x/x_c>1$, the system is at the quantum critical line depicted
by the dash line and shown by the vertical arrow. The critical line
is characterized by the FC state with finite superconducting order
parameter $\kappa$. At $T_c=0$, the order parameter $\kappa$ is
destroyed, the system undergoes the first order phase transition and
exhibits the NFL behavior at $T>0$.}\label{fig1}
\end{figure}

Schematic phase diagram of the system which is driven to FC by
variation of $x$ is reported in Fig. \ref{fig1}.  Upon approaching
the critical density $x_c$ the system remains in LFL region at
sufficiently low temperatures \cite{khodb,obz}, that is shown by the
shadow area. At QCP $x_c$ shown by the arrow in Fig. \ref{fig1}, the
system demonstrates the NFL behavior down to the lowest
temperatures. Beyond QCP at finite temperatures the behavior is
remaining the NFL one and is determined by the
temperature-independent entropy $S_0$ \cite{yakov}. In that case at
$T\to 0$, the system is approaching a quantum critical line (shown
by the vertical arrow and the dashed line in Fig. \ref{fig1}) rather
than a quantum critical point. Upon reaching the quantum critical
line from the above at $T\to0$ the system undergoes the first order
quantum phase transition, which is FCQPT taking place at $T_c=0$.

At $T>0$ the NFL state above the critical line, see Fig. \ref{fig1},
is strongly degenerated, therefore it is captured by the other
states such as superconducting (for example, by the superconducting
state in $\rm CeCoIn_5$ \cite{shag1,shag2,yakov}) or by AF state
(e.g. AF one in $\rm YbRh_2Si_2$ \cite{plaq}) lifting the
degeneracy. The application of magnetic field $B>B_{c0}$ restores
the LFL behavior, where $B_{c0}$ is a critical magnetic field, such
that at $B>B_{c0}$ the system is driven towards its Landau Fermi
liquid (LFL) regime \cite{shag2}. In some cases, for example in HF
metal $\rm CeRu_2Si_2$, $B_{c0}=0$, see e.g. \cite{takah}, while in
$\rm YbRh_2Si_2$, $B_{c0}\simeq 0.06$ T \cite{geg}. In our simple
model $B_{c0}$ is taken as a parameter.

\begin{figure}[!ht]
\begin{center}
\vspace*{-0.5cm}
\includegraphics [width=0.44\textwidth]{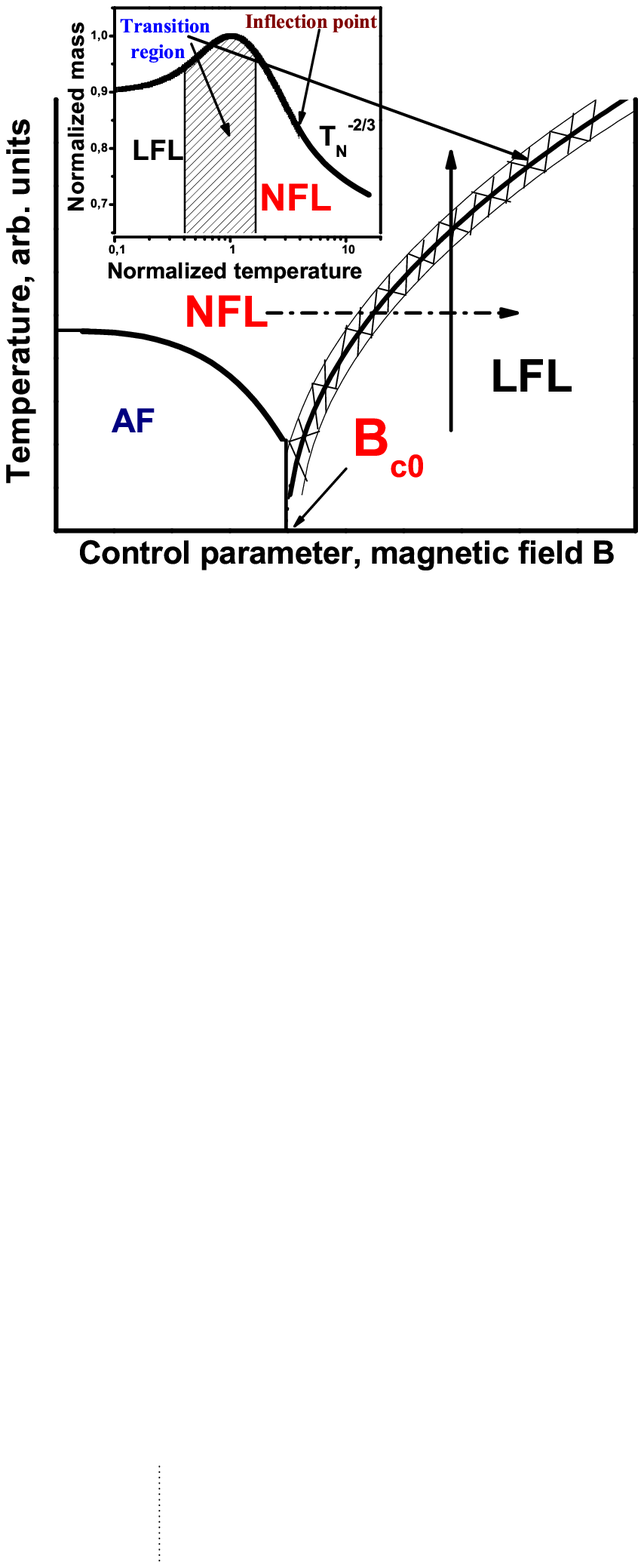}
\vspace*{-11.8cm}
\end{center}
\caption{ Schematic phase diagram of the HF metal $\rm YbRh_2Si_2$.
$B_{c0}$ is magnetic field at which the effective mass diverges.
$\rm {AF}$ denotes antiferromagnetic (AF) state. At $B<B_{c0}$ the
system is in AF state. The vertical arrow shows the transition from
the LFL regime to the NFL one at fixed $B$ along $T$ with $M^*$
depending on $T$. The dash-dot horizontal arrow illustrates the
system moving from NFL regime to LFL one along $B$ at fixed $T$.
The inset shows a schematic plot of the scaling behavior of the
normalized effective mass versus the normalized temperature.
Transition regime, where $M^*_N$ reaches its maximum value $M^*_M$
at $T=T_M$, is shown by the hatched area both in the main panel and
in the inset. The arrows mark the position of inflection point in
$M^*_N$ and the transition region.}\label{PHD}
\end{figure}

Schematic phase diagram of the HF metal $\rm YbRh_2Si_2$ is reported
in Fig. \ref{PHD}. Magnetic field $B$ is taken as the control
parameter. The FC state and the region lying at $x/x_c\geq 1$, see
Fig. \ref{fig1}, can be captured by the superconducting,
ferromagnetic, antiferromagnetic (AF) etc. states lifting the
degeneracy \cite{obz,khodb}. Since we consider the HF metal $\rm
YbRh_2Si_2$ the AF state takes place \cite{geg} as shown in Fig.
\ref{PHD}. As seen from Fig. \ref{PHD}, at elevated temperatures and
fixed magnetic field the NFL regime occurs, while rising $B$ again
drives the system from NFL region to LFL one. Below we consider the
transition region when at rising $B$ the system moves from NFL
regime to LFL one along the dash-dot horizontal arrow, and at
elevated $T$ it moves from LFL regime to NFL one along the solid
vertical arrow.

To explore a scaling behavior of $M^*$, we write the quasiparticle
distribution function as $n_1({\bf p})=n({\bf p},T)-n({\bf p})$,
with $n({\bf p})$ is the step function, and Eq. \eqref{LANDM} then
becomes
\begin{equation}
\frac{1}{M^*(T)}=\frac{1}{M^*}+\int\frac{{\bf p}_F{\bf
p_1}}{p_F^3}F({\bf p_F},{\bf p}_1)\frac{\partial n_1(p_1,T)}
{\partial p_1}\frac{d{\bf p}_1}{(2\pi) ^3}. \label{LF1}
\end{equation}
At QCP the effective mass $M^*$ diverges and Eq. \eqref{LF1} becomes
homogeneous determining $M^*$ as a function of temperature
\begin{equation}M^*(T)\propto T^{-2/3},\label{LTT}
\end{equation}
while the system exhibits the NFL behavior \cite{ckz,obz}. If the
system is located before QCP, $M^*$ is finite, at low temperatures
the system demonstrates the LFL behavior that is $M^*(T)\simeq
M^*+a_1T^2$, with $a_1$ is a constant, see the inset to Fig.
\ref{PHD}. Obviously, the LFL behavior takes place when the second
term on the right hand side of Eq. \eqref{LF1} is small in
comparison with the first one. Then, at rising temperatures the
system enters the transition regime: $M^*$ grows, reaching its
maximum $M^*_M$ at $T=T_M$, with subsequent diminishing. Near
temperatures $T\geq T_M$ the last "traces" of LFL regime disappear,
the second term starts to dominate, and again Eq. \eqref{LF1}
becomes homogeneous, and the NFL behavior restores, manifesting
itself in decreasing $M^*$ as $T^{-2/3}$. When the system is near
QCP, it turns out that the solution of Eq. \eqref{LF1} $M^*(T)$ can
be well approximated by a simple universal interpolating function
\cite{obz,ckz,shag2}. The interpolation occurs between the LFL
($M^*\simeq M^*+a_1T^2$) and NFL ($M^*\propto T^{-2/3}$) regimes
thus describing the above crossover \cite{ckz,obz}. Introducing the
dimensionless variable $y=T_N=T/T_M$, we obtain the desired
expression \begin{equation}M^*_N(y)\approx
c_0\frac{1+c_1y^2}{1+c_2y^{8/3}}. \label{UN2}
\end{equation}
Here $M^*_N=M^*/M^*_M$ is the normalized effective mass,
$c_0=(1+c_2)/(1+c_1)$, $c_1$ and $c_2$ are fitting parameters,
parameterizing the Landau amplitude.

The inset to Fig. \ref{PHD} demonstrates the scaling behavior of the
normalized effective mass $M^*_N=M^*/M^*_M$ versus normalized
temperature $T_N=T/T_M$, where $M^*_M$ is the maximum value that
$M^*$ reaches at $T=T_M$. At $T\ll T_M$ the LFL regime takes place.
At $T\gg T_M$ the $T^{-2/3}$ regime takes place. This is marked as
NFL one since the effective mass depends strongly on temperature.
The temperature region $T\simeq T_M$ signifies the transition
between the LFL regime with almost constant effective mass and NFL
behavior, given by $T^{-2/3}$ dependence. Thus temperatures $T\sim
T_M$ can be regarded as the transition region between LFL and NFL
regimes. The transition temperatures are not really a phase
transition. These necessarily are broad, very much depending on the
criteria for determination of the point of such a transition, as it
is seen from the inset to Fig. \ref{PHD}. As usually, the transition
temperature is extracted from the temperature dependence of charge
transport, for example, from the resistivity $\rho(T)=\rho_0+AT^2$
with $\rho_0$ is the residual resistivity and $A$ is the LFL
coefficient. The crossover takes place at temperatures where the
resistance starts to deviate from the LFL $T^2$ behavior. Obviously,
the measure of the deviation from the LFL $T^2$ behavior cannot be
defined unambiguously. Therefore, different measures produce
different results.

It is possible to transport Eq. \eqref{LF1} to the case of the
application of magnetic fields \cite{ckz,obz,shag2}.  The
application of magnetic field restores the LFL behavior so that
$M^*_M$ depends on $B$ as
\begin{equation}\label{LFLB}
    M^*_M\propto (B-B_{c0})^{-2/3},
\end{equation} while
\begin{equation}\label{LFLT}T_M\propto \mu_B(B-B_{c0}),\end{equation}
where $\mu_B$ is the Bohr magneton \cite{ckz,shag2,obz}. Employing
Eqs. \eqref{LFLB} and \eqref{LFLT} to calculate $M^*_M$ and $T_M$,
we conclude that Eq. \eqref{UN2} is valid to describe the normalized
effective mass in external fixed magnetic fields with
$y=T/(B-B_{c0})$. On the other hand, Eq. \eqref{UN2} is valid when
the applied magnetic field becomes a variable, while temperature is
fixed $T=T_f$. In that case, as seen from Eqs. \eqref{LTT},
\eqref{UN2} and\eqref{LFLB}, it is convenient to rewrite both the
variable as $y=(B-B_{c0})/T_f$, and Eq. \eqref{LFLT} as
\begin{equation}\label{LFLf}\mu_B(B_M-B_{c0})\propto T_f.\end{equation}

It follows from Eq. \eqref{UN2} that in contrast to the Landau
paradigm of quasiparticles the effective mass strongly depends on
$T$ and $B$. As we will see it is this dependence that forms the NFL
behavior. It follows also from Eq. \eqref{UN2} that a scaling
behavior of $M^*$ near QCP is determined by the absence of
appropriate external physical scales to measure the effective mass
and temperature. At fixed magnetic fields, the characteristic scales
of temperature and of the function $M^*(T,B)$ are defined by both
$T_M$ and $M^*_M$ respectively. At fixed temperatures, the
characteristic scales are $(B_M-B_{c0})$ and $M^*_M$. It follows
from Eqs. \eqref{LFLB} and \eqref{LFLT} that at fixed magnetic
fields, $T_M\to0$, and $M^*_M\to\infty$, and the width of the
transition region shrinks to zero as $B\to B_{c0}$ when these are
measured in the external scales. In the same way, it follows from
Eqs. \eqref{LTT} and \eqref{LFLf} that at fixed temperatures,
$(B_M-B_{c0})\to0$, and $M^*_M\to\infty$, and the width of the
transition region shrinks to zero as $T_f\to0$. Thus, the
application of the external scales obscure the scaling behavior of
the effective mass and thermodynamic and transport properties.

\begin{figure} [! ht]
\begin{center}
\vspace*{-0.5cm}
\includegraphics [width=0.49\textwidth]{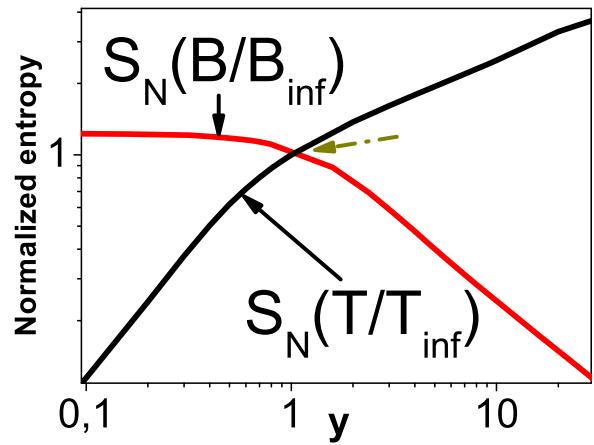}
\end{center}
\vspace*{-1.2cm} \caption{The normalized entropy $S_N(B/B_{inf})$
versus $y=B/B_{inf}$ and the normalized entropy $S_N(T/T_{inf})$
versus $y=T/T_{inf}$ calculated at fixed temperature and magnetic
field, correspondingly, are represented by the solid lines and shown
by the arrows. The inflection point is depicted by the dash-dot
arrow.}\label{STB}
\end{figure}

A few remarks are in order here. As we shall see, magnetic field
dependencies of the effective mass or of other observable like the
longitudinal magnetoresistance do not have "peculiar points" like
maximum. The normalization are to be performed in the other points
like the inflection point at $T=T_{inf}$ (or at $B=B_{inf}$) shown
in the inset to Fig. \ref{PHD} by the arrow. Such a normalization is
possible since it is established on the internal scales,
$T_{inf}\propto T_M\propto(B-B_{c0})$.

In what follows, we compute the effective mass and employ Eq.
\eqref{UN2} for estimations of considered values. To compute the
effective mass $M^*(T,B)$, we solve Eq. \eqref{LF1} with special
form of Landau interaction amplitude, see Refs. \cite{ckz,obz} for
details. Choice of the amplitude is dictated by the fact that the
system has to be at QCP, which means that first two $p$-derivatives
of the single-particle spectrum $\varepsilon({\bf p})$ should equal
zero. Since first derivative is proportional to the reciprocal
quasiparticle effective mass $1/M^*$, its zero just signifies QCP of
FCQPT. Zeros of two subsequent derivatives mean that the spectrum
$\varepsilon({\bf p})$ has an inflection point at $p_F$ so that the
lowest term of its Taylor expansion is proportional to $(p-p_F)^3$
\cite{ckz}. After solution of Eq. \eqref{LF1}, the obtained spectrum
had been used to calculate the entropy $S(B,T)$, which, in turn, had
been recalculated to the effective mass $M^*(T,B)$ by virtue of
well-known LFL relation $M^*(T,B)=S(T,B)/T$. Our calculations of the
normalized entropy as a function of the normalized magnetic field
$B/B_{inf}=y$ and as a function of the normalized temperature
$y=T/T_{inf}$ are reported in Fig. \ref{STB}. Here $T_{inf}$ and
$B_{inf}$ are the corresponding inflection points in function $S$.
We normalize the entropy by its value at the inflection point
$S_N(y)=S(y)/S(1)$. As seen form Fig. \ref{STB}, our calculations
corroborate the scaling behavior of the normalized entropy, that is
the curves at different temperatures and magnetic fields merge into
single one in terms of the variable $y$. The inflection point
$T_{inf}$ in $S(T)$ makes $M^*(T,B)$ have its maximum as a function
of $T$, while $M^*(T,B)$ versus $B$ has no maximum. We note that our
calculations of the entropy confirm the validity of Eq. \eqref{UN2}
and the scaling behavior of the normalized effective mass.

Exciting measurements of $C/T\propto M^*$ on samples of the new
generation of $\rm YbRh_2Si_2$ in different magnetic fields $B$ up
to 1.5 T \cite{oes} allow us to identify the scaling behavior of the
effective mass $M^*$ and observe the different regimes of $M^*$
behavior such as the LFL regime, transition region from LFL to NFL
regimes, and the NFL regime itself. A maximum structure in
$C/T\propto M^*_M$ at $T_M$ appears under the application of
magnetic field $B$ and $T_M$ shifts to higher $T$ as $B$ is
increased. The value of $C/T=\gamma_0$ is saturated towards lower
temperatures decreasing at elevated magnetic field, where $\gamma_0$
is the Sommerfeld coefficient \cite{oes}.

The transition region corresponds to the temperatures where the
vertical arrow in the main panel of Fig. \ref{PHD} crosses the
hatched area. The width of the region, being proportional to
$T_M\propto (B-B_{c0})$ shrinks,  $T_M$ moves to zero temperature
and $\gamma_0\propto M^*$ increases as $B\to B_{c0}$. These
observations are in accord with the facts \cite{oes}.

\begin{figure} [! ht]
\begin{center}
\vspace*{-0.2cm}
\includegraphics [width=0.49\textwidth]{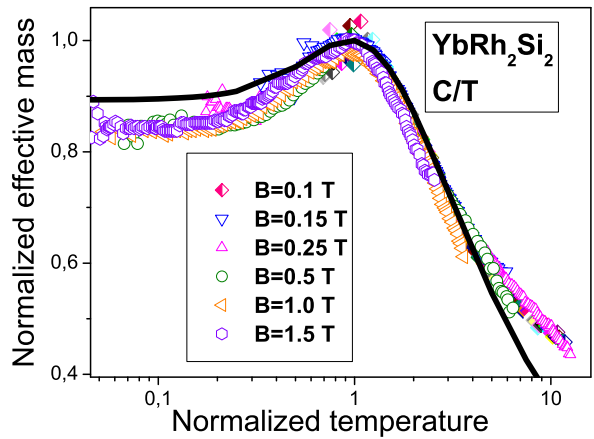}
\end{center}
\vspace*{-0.3cm} \caption{The normalized effective mass $M^*_N$
extracted from the measurements of the specific heat $C/T$ on $\rm
YbRh_2Si_2$ in magnetic fields $B$ \cite{oes} listed in the legend.
Our calculations are depicted by the solid curve tracing the scaling
behavior of $M^*_N$.}\label{fig2}
\end{figure}

To obtain the normalized effective mass $M^*_N$, the maximum
structure in $C/T$ was used to normalize $C/T$, and $T$ was
normalized by $T_M$. In Fig. \ref{fig2} $M^*_N$ as a function of
normalized temperature $T_N$ is shown by geometrical figures, our
calculations are shown by the solid line. Figure \ref{fig2} reveals
the scaling behavior of the normalized experimental curves - the
scaled curves at different magnetic fields $B$ merge into a single
one in terms of the normalized variable $y=T/T_M$. As seen, the
normalized mass $M^*_N$ extracted from the measurements is not a
constant, as would be for  LFL. The two regimes (the LFL regime and
NFL one) separated by the transition region, as depicted by the
hatched area in the inset to Fig. \ref{PHD}, are clearly seen in
Fig. \ref{fig2} illuminating good agreement between the theory and
facts. It is worthy of note that the normalization procedure allows
us to construct the scaled function $C/T$ extracted from the facts
in wide range variation of the normalized temperature. Indeed, it
integrates measurements of $C/T$ taken at the application of
different magnetic fields into unique function which demonstrates
the scaling behavior over three decades in normalized temperature as
seen from Fig. \ref{fig2}.

\begin{figure} [! ht]
\begin{center}
\vspace*{-0.5cm}
\includegraphics [width=0.49\textwidth]{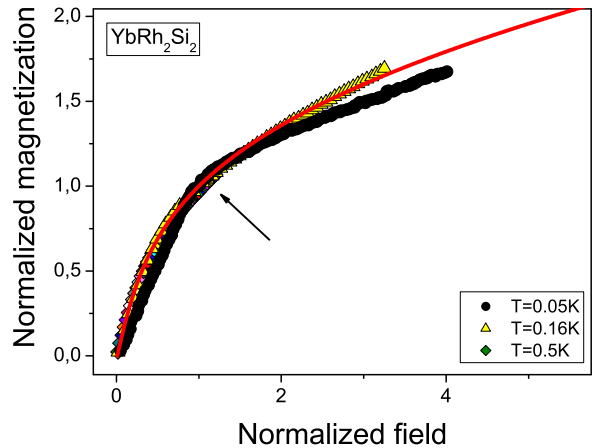}
\end{center}
\vspace*{-0.8cm} \caption{The field dependencies of the normalized
magnetization $M$ collected at different temperatures shown at right
bottom corner are extracted from measurements collected on $\rm{
YbRu_2Si_2}$ \cite{steg}. The kink (shown by the arrow) is clearly
seen at the normalized field $B_N=B/B_k\simeq 1$. The solid curve
represents our calculations.}\label{fig3}
\end{figure}

Consider now the magnetization $M$ as a function of magnetic field
$B$ at fixed temperature $T=T_f$
\begin{equation}\label{CHIB}
M(B,T)=\int_0^B \chi(b,T)db,
\end{equation}
where the magnetic susceptibility $\chi$ is given by \cite{land}
\begin{equation}\label{CHI}
\chi(B,T)=\frac{\beta M^*(B,T)}{1+F_0^a}.
\end{equation}
Here, $\beta$ is a constant and $F_0^a$ is the Landau amplitude
related to the exchange interaction \cite{land}. In the case of
strongly correlated systems $F_0^a\geq -0.9$ \cite{pfit,pfit1}.
Therefore, as seen from Eq. \eqref{CHI}, due to the normalization
the coefficients $\beta$ and $(1+F_0^a)$ drops out from the result,
and $\chi\propto M^*$.

One might suppose that $F_0^a$ can strongly depend on $B$. This is
not the case, since the Kadowaki-Woods ratio is conserved
\cite{kadw,geg1,kwz}, $A(B)/\gamma_0^2(B)\propto
A(B)/\chi^2(B)\propto const$, we have $\gamma_0\propto M^*\propto
\chi$. Here $A$ is the coefficient in the $T^2$ dependence of
resistivity $\rho$.

Our calculations show that the magnetization exhibits a kink at some
magnetic field $B=B_k$. The experimental magnetization demonstrates
the same behavior \cite{steg}. We use $B_k$ and $M(B_k)$ to
normalize $B$ and $M$ respectively. The normalized magnetization
$M(B)/M(B_k)$ extracted from facts \cite{steg} depicted by the
geometrical figures and calculated magnetization shown by the solid
line are reported in Fig. \ref{fig3}. As seen, the scaled data at
different $T_f$ merge into a single one in terms of the normalized
variable $y=B/T_k$. It is also seen, that these exhibit energy
scales separated by kink at the normalized magnetic field
$B_N=B/B_k=1$. The kink is a crossover point from the fast to slow
growth of $M$ at rising magnetic field. It is seen from Fig.
\ref{fig3}, that our calculations are in good agreement with the
facts, and all the data exhibit the kink (shown by the arrow) at
$B_N\simeq 1$ taking place as soon as the system enters the
transition region corresponding to the magnetic fields where the
horizontal dash-dot arrow in the main panel of Fig. \ref{PHD}
crosses the hatched area. Indeed, as seen from Fig. \ref{fig3}, at
lower magnetic fields $M$ is a linear function of $B$ since $M^*$ is
approximately independent of $B$. Then, it follows from Eqs.
\eqref{UN2} and \eqref{LFLB} that at elevated magnetic fields $M^*$
becomes a diminishing function of $B$ and generates the kink in
$M(B)$ separating the energy scales discovered in Refs.
\cite{steg1,steg}. Then, as seen from Eq. \eqref{LFLf} the magnetic
field $B_k$ at which the kink appears, $B_k\simeq B_M\propto T_f$,
shifts to lower $B$ as $T_f$ is decreased. This observation is in
accord with facts \cite{steg1,steg}.

\begin{figure} [! ht]
\begin{center}
\vspace*{-0.5cm}
\includegraphics [width=0.49\textwidth]{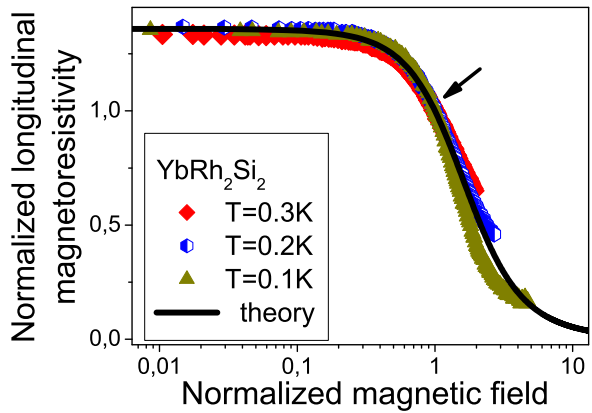}
\end{center}
\vspace*{-0.8cm} \caption{Magnetic field dependence of the
normalized magnetoresistance $\rho_N$ versus normalized magnetic
field. $\rho_N$ was extracted from LMR of $\rm YbRh_2Si_2$ at
different temperatures \cite{steg1,steg} listed in the legend. The
inflection point is shown by the arrow, and the solid line
represents our calculations.}\label{fig4}
\end{figure}

Consider a longitudinal magnetoresistance (LMR)
$\rho(B,T)=\rho_0+AT^2$ as a function of $B$ at fixed $T_f$. In that
case, the classical contribution to LMR due to orbital motion of
carriers induced by the Lorentz force is small, while the
Kadowaki-Woods relation \cite{kadw,geg1,kwz}, $K=A/\gamma_0^2\propto
A/\chi^2=const$, allows us to employ $M^*$ to construct the
coefficient $A$ \cite{pla3}, since $\gamma_0\propto\chi\propto M^*$.
As a result, $\rho(B,T)-\rho_0\propto(M^*)^2$. Fig. \ref{fig4}
reports the normalized magnetoresistance
\begin{equation}\label{rn}
\rho_N(y)=\frac{\rho(y)-\rho_0}{\rho_{inf}}\propto (M_N^*(y))^2
\end{equation}
versus normalized magnetic field $y=B/B_{inf}$ at different
temperatures, shown in the legend. Here $\rho_{inf}$ and $B_{inf}$
are LMR and magnetic field respectively taken at the inflection
point marked by the arrow in Fig. \ref{fig4}. Both theoretical
(shown by the solid line) and experimental (marked by the
geometrical figures) curves have been normalized by their inflection
points, which also reveals the scaling behavior - the scaled curves
at different temperatures merge into single one as a function of the
variable $y$ and show the scaling behavior over three decades in the
normalized magnetic field. The transition region at which LMR starts
to decrease is shown in the inset to Fig. \ref{PHD} by the hatched
area. Obviously, as seen from Eq. \eqref{LFLf}, the width of the
transition region being proportional to $B_M\simeq B_{inf}\propto
T_f$ decreases as the temperature $T_f$ is lowered. In the same way,
the inflection point of LMR, generated by the inflection point of
$M^*$ shown in the inset to Fig. \ref{PHD} by the arrow, shifts to
lower $B$ as $T_f$ is decreased. All these observations are in
excellent agreement with the facts \cite{steg1,steg}.

The evolution of the derivative of magnetic entropy $dS(B,T)/dB$ as
a function of magnetic field $B$ at fixed temperature $T_f$ is of
great importance since it allows us to study the scaling behavior of
the derivative of the effective mass $TdM^*(B,T)/dB\propto
dS(B,T)/dB$. While the scaling properties of the effective mass
$M^*(B,T)$ can be analyzed via LMR, see Fig. \ref{fig4}.

\begin{figure} [! ht]
\begin{center}
\includegraphics [width=0.47\textwidth]{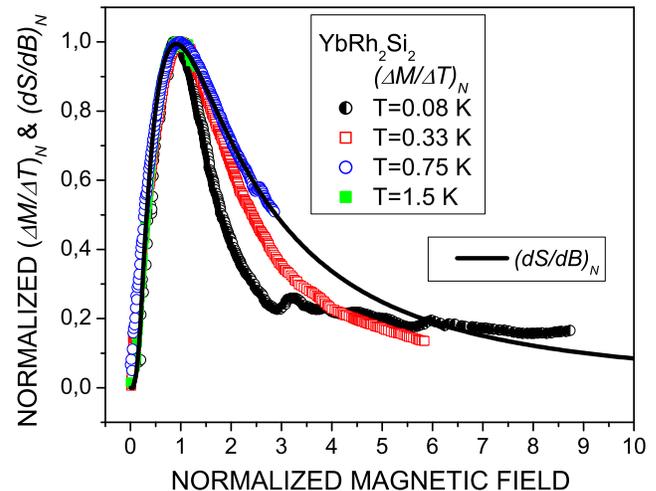}
\vspace*{-0.6cm}
\end{center}
\caption{Normalized magnetization difference divided by temperature
increment $(\Delta M/\Delta T)_N$ versus normalized magnetic field
at fixed temperatures listed in the legend is extracted from the
facts collected  on $\rm YbRh_2Si_2$ \cite{geg}. Our calculation of
the normalized derivative $(dS/dB)_N\simeq (\Delta M/\Delta T)_N$
versus normalized magnetic field is shown by the solid line.}
\label{fig5}
\end{figure}

As seen from from Eqs. \eqref{UN2} and \eqref{LFLf}, at $y\leq 1$
the derivative $-dM_N(y)/dy\propto y$ with
$y=(B-B_{c0})/(B_{inf}-B_{c0})\propto (B-B_{c0})/T_f$. We note that
the effective mass as a function of $B$ does not have the maximum.
At elevated $y$ the derivative $-dM_N(y)/dy$ possesses a maximum at
the inflection point and then becomes a diminishing function of $y$.
Upon using the variable $y=(B-B_{c0})/T_f$, we conclude that at
decreasing temperatures, the leading edge of the function
$-dS/dB\propto -TdM^*/dB$ becomes steeper and its maximum at
$(B_{inf}-B_{c0})\propto T_f$ is higher. These observations are in
quantitative agreement with striking measurements of the
magnetization difference divided by temperature increment, $-\Delta
M/\Delta T$, as a function of magnetic field at fixed temperatures
$T_f$ collected on $\rm YbRh_2Si_2$ \cite{geg}. We note that
according to the well-know thermodynamic equality $dM/dT=dS/dB$, and
$\Delta M/\Delta T\simeq dS/dB$. To carry out a quantitative
analysis of the scaling behavior of $-dM^*(B,T)/dB$, we calculate
the normalized entropy $S$ shown in Fig. \ref{STB} as a function of
$B/B_{inf}$ at fixed temperature $T_f$. Fig. \ref{fig5} reports the
normalized $(dS/dB)_N$ as a function of the normalized magnetic
field. The scaled function $(dS/dB)_N$ is obtained by normalizing
$(-dS/dB)$ by its maximum taking place at $B_M$, and the field $B$
is scaled by $B_M$. The measurements of $-\Delta M/\Delta T$ are
normalized in the same way and depicted in Fig. \ref{fig5} as
$(\Delta M/\Delta T)_N$ versus normalized field. It is seen from
Fig. \ref{fig5} that our calculations shown by the solid line are in
good agreement with the facts and the scaled functions $(\Delta
M/\Delta T)_N$ extracted from the facts show the scaling behavior in
wide range variation of the normalized magnetic field $B/B_M$.
\begin{figure} [! ht]
\begin{center}
\vspace*{-0.6cm}
\includegraphics [width=0.49\textwidth]{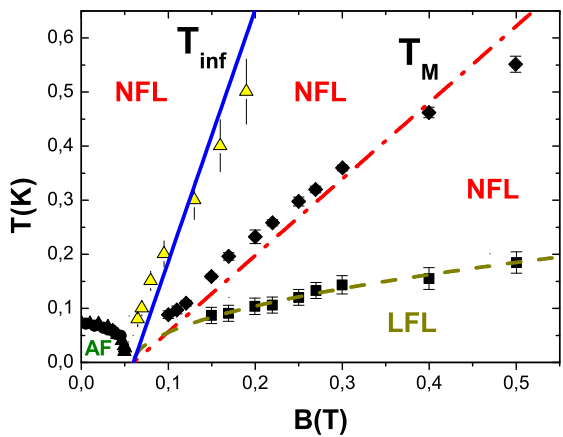}
\end{center}
\vspace*{-0.8cm} \caption{Temperature versus magnetic field $T-B$
phase diagram for $\rm YbRh_2Si_2$. Solid circles represent the
boundary between AF and NFL states. The solid squares  denote the
boundary of the NFL and LFL regime \cite{steg1,steg,geg1} shown by
the dotted line which is approximated by $\sqrt{B-B_{c0}}$
\cite{obz}. Diamonds mark the maximums $T_M$ of $C/T$ \cite{oes}
shown in Fig. \ref{fig2}. The dash-dot line is approximated by
$T_M\propto a(B-B_{c0})$, $a$ is a fitting parameter, see Eq.
\eqref{LFLT}. Triangles along the solid line denote $T_{inf}$ in LMR
\cite{steg1,steg} sown in Fig. \ref{fig5}, the solid line represents
the function $T_{inf}\propto b(B-B_{c0})$, $b$ is a fitting
parameter, see Eq. \eqref{LFLf}.}\label{fig6}
\end{figure}

Fig. \ref{fig6} reports $T_{inf}$ and $T_M$ versus $B$ depicted by
the solid and dash-dotted lines, respectively. The boundary between
the NFL and LFL regimes is shown by the dashed line, and AF marks
the antiferromagnetic state. The corresponding data are taken from
Ref. \cite{steg1,steg,oes,geg1}. It is seen that our calculations
are in good agreement with the facts. In Fig. \ref{fig6}, the solid
and dash-dotted lines corresponding to the functions $T_{inf}$ and
$T_M$, respectively, represent the positions of the kinks separating
the energy scales in $C$ and $M$ reported in Ref. \cite{steg1,steg}.
It is seen that our calculations are in accord with facts, and we
conclude that the energy scales are reproduced by Eqs. \eqref{LFLT}
and \eqref{LFLf} and related to the peculiar points $T_{inf}$ and
$T_M$ of the normalized effective mass $M^*_N$ which are shown by
the arrows in the inset to Fig. \ref{PHD}.

At $B\to B_{c0}$ both $T_{inf}\to 0$ and $T_{M}\to 0$, thus the LFL
and the transition regimes of both $C/T$ and $M$ as well as these of
LMR and the magnetic entropy are shifted to very low temperatures.
Therefore due to experimental difficulties these regimes cannot be
often observed in experiments on HF metals. As it is seen from Figs.
\ref{fig2}, \ref{fig3}, \ref{fig4}, \ref{fig5} and \ref{fig6}, the
normalization allows us to construct the unique scaled thermodynamic
and transport functions extracted from the experimental facts in
wide range of the variation of the scaled variable $y$. As seen from
the mentioned Figures, the constructed normalized thermodynamic and
transport functions show the scaling behavior over three decades in
the normalized variable.

In summary, we have analyzed the non-Fermi liquid behavior of the
heavy fermion metals, and showed that extended quasiparticles
paradigm is strongly valid, while the dependence of the effective
mass on temperature, number density and applied magnetic fields
gives rise to the NFL behavior. We have demonstrated that our
theoretical study of the heat capacity, magnetization, longitudinal
magnetoresistance and magnetic entropy are in good agreement with
the outstanding recent facts collected on the HF metal $\rm
YbRh_2Si_2$. Our normalization procedure has allowed us to construct
the scaled thermodynamic and transport properties in wide range of
the variation of the scaled variable. For $\rm YbRh_2Si_2$ the
constructed thermodynamic and transport functions show the scaling
behavior over three decades in the normalized variable. The energy
scales in these functions are also explained.

This work was supported in part by the grants: RFBR No. 09-02-00056
and the Hebrew University Intramural Funds. V.R.S. is grateful to
the Lady Davis Foundation for supporting his visit to the Hebrew
University of Jerusalem.

\end{document}